\documentclass[UKenglish,a4paper,cleveref,capitalize,nameinlink,noabbrev,svgnames]{lipics-v2021}
\usepackage[shortcuts]{extdash}
\usepackage{xfrac}
\usepackage{mathtools}
\usepackage{amsthm}
\usepackage{amssymb}
\usepackage{braket}
\usepackage{standalone}

\usepackage{ragged2e}
\usepackage[inline]{enumitem}
\usepackage[onelanguage,linesnumbered,vlined,algoruled,procnumbered]{algorithm2e}

\usepackage{xparse}
\usepackage{tabularray}

\usepackage{marginnote}
\usepackage{todonotes}

\makeatletter
\let\@scshape@orig=\scschape
\newcommand*\IfBoldTF{\ifx\f@series\m@series@bold\expandafter\@firstoftwo\else\expandafter\@secondoftwo\fi}
\newcommand*\m@series@bold{bx}
\def\scshape{\IfBoldTF{\expandafter\MakeUppercase}{\@scshape@orig}}
\def\textsc#1{\ifmmode\text{\IfBoldTF{\MakeUppercase{#1}}{\scshape#1}}\else{\IfBoldTF{\MakeUppercase{#1}}{\scshape#1}}\fi}
\makeatother

\NewDocumentCommand\gbtodo{s o m}{\todo[author=B,prepend,caption={\strong{B}},size=\footnotesize,color=yellow,noauthor\IfValueT{#2}{,#2}\IfBooleanT{#1}{,inline}]{#3}}

\usepackage{tikz}
\hypersetup{colorlinks=true,pdfusetitle,allcolors=blue!50!black}
\newcommand\crefw[2][]{\hyperref[#2]{#1~\labelcref{#2}}}
\newcommand\refw[2]{\hyperref[#2]{#1~\labelcref{#2}}}

\NewCommandCopy{\legacyunderscore}{\_}
\renewcommand{\_}{\ifincsname_\else\legacyunderscore\fi}
\DontPrintSemicolon
\SetProcNameSty{texttt}
\SetKwFunction{consecutive}{{Consecutive}$_n$}
\SetKwFunction{isprefbinaryseq}{{isPrefBinarySeq}$_n$}
\SetKwFunction{isFactBinSeq}{{isFactBinarySeq}$_n$}
\SetKwFunction{issameprefix}{{isSamePrefix}$_n$}
\SetKwFunction{issamesuffix}{{isSameSuffix}$_n$}
\SetKwFunction{issameblock}{{isSameBlock}$_n$}
\SetKwFunction{readrelative}{readRelative}
\SetKwFunction{read}{read}
\SetKwFunction{writecell}{write}
\SetKw{accept}{accept}
\SetKw{reject}{reject}
\SetKw{Break}{break}
\SetKw{KwAnd}{and}
\SetKw{KwOr}{or}
\SetKwData{swap}{doswap}
\SetKwData{true}{true}
\SetKwData{false}{false}

\Crefname{algorithm}{Procedure}{Procedures}

\newtheorem{fact}{Fact}

\def\strongsty{\bfseries\mathversion{bold}}
\def\strong#1{{\strongsty#1}}
\def\defemsty{\color{DarkGreen!85!black}\ifmmode\else\em\fi}
\def\defem#1{{\defemsty#1}}

\newcommand{\Nat}{\ensuremath{\mathbb{N}}\xspace}
\newcommand{\emptyword}{\ensuremath{\varepsilon}\xspace}
\let\ew=\emptyword
\let\dollar=\$
\def\ssep{\ensuremath{\sharp}\xspace}
\def\dsep{\ensuremath{\dollar}\xspace}

\newcommand\s{\text{s}\xspace}
\newcommand\ow{\textsc1}
\newcommand\tw{\textsc2}
\newcommand\fa{\textsc{fa}}
\newcommand\twfa{\textsc{2}\fa}
\newcommand\dfa{\textsc{d}\fa}
\newcommand\nfa{\textsc{n}\fa}
\newcommand\twdfa{\ensuremath{\tw\dfa}\xspace}
\newcommand\twnfa{\ensuremath{\tw\nfa}\xspace}
\newcommand\owdfa{\ensuremath{\ow\dfa}\xspace}
\newcommand\ownfa{\ensuremath{\ow\nfa}\xspace}

\newcommand{\la}{\textsc{1\=/la}\xspace}
\newcommand{\dla}{\textsc{d}\la}
\newcommand\smallplus{\ensuremath{\raisebox{.40\height}{\scalebox{.6}{\text{+}}}}}
\NewDocumentCommand{\twdfacg}{o}{\ensuremath{{\twdfa\IfValueTF{#1}{#1}{\!}\smallplus\mathsf{cg}}}\xspace}
\NewDocumentCommand{\twnfacg}{o}{\ensuremath{{\twnfa\IfValueTF{#1}{#1}{\!}\smallplus\mathsf{cg}}}\xspace}

\newcommand\twdfas{\twdfa\s}
\newcommand\twnfas{\twnfa\s}
\newcommand\owdfas{\owdfa\s}
\newcommand\ownfas{\ownfa\s}
\newcommand\las{\la\s}
\newcommand\dlas{\dla\s}
\newcommand{\twdfacgs}{\twdfacg[\s]}
\newcommand{\twnfacgs}{\twnfacg[\s]}

\newcommand{\machine}[1]{\ensuremath{\mathcal{#1}}\xspace}
\newcommand{\lang}[1]{\ensuremath{\mathsf{#1}}\xspace}
\newcommand{\norm}[1]{{\pi_0\left(#1\right)}}
\newcommand{\rend}{\ensuremath{\triangleleft}\xspace}
\newcommand{\lend}{\ensuremath{\triangleright}\xspace}
\newcommand{\cfg}[3]{\ensuremath{\ifx&#1&\else{#1\cdot}\fi#2\ifx&#3&\else{\cdot#3}\fi}}

\newcommand{\langof}[1]{\ensuremath{\mathcal{L}(#1)}}
\newcommand{\pref}[1]{\ensuremath{\mathrm{Pref}(#1)}}
\newcommand{\suff}[1]{\ensuremath{\mathrm{Suff}(#1)}}
\newcommand{\factor}[1]{\ensuremath{\mathrm{Fact}(#1)}}

\newcommand{\bin}[2][n]{\ensuremath{b_{#1}\ifx*#2\else(#2)\fi}}
\newcommand{\fbs}[1][n]{\ensuremath{\mathsf{FullBinSeq}_{#1}}\xspace}
\newcommand{\ifbs}[1][n]{\ensuremath{\mathsf{IterSuffFullBinSeq}_{#1}}\xspace}
\newcommand{\sameprefix}[1]{\ensuremath{\mathsf{SamePrefix}_{#1}}}

\DeclareRobustCommand{\bigO}{\ensuremath{\text{\usefont{OMS}{cmsy}{m}{n}O}}}
\newcommand{\primorial}[1]{\ensuremath{#1\#}}
\newcommand{\compl}[1]{\ensuremath{\overline{#1}}}
\newcommand{\card}[1]{\ensuremath{\lvert{#1}\rvert}}

\newcommand*\ie{\emph{i.e.}\xspace}
\newcommand*\cf{\emph{c.f.}\xspace}
\newcommand*\eg[1][e]{\emph{#1.g.}\xspace}
\newcommand*\resp{{resp.}\xspace}
\newcommand*\ifof{if and only if\xspace}
\newcommand*\Wlog[1][W]{#1ithout loss of generality\xspace}

\xspaceaddexceptions{\s \xspace}

\newif\ifshort\shortfalse

\nolinenumbers
\hideLIPIcs
\def\subjclassHeading{}
\def\keywordsHeading{\vspace{-2\baselineskip}}

\title{Nondeterminism makes unary 1-limited automata concise}
\author{Bruno Guillon}{Université Clermont Auvergne, Clermont Auvergne INP, LIMOS, CNRS, F\=/63000 Clermont-Ferrand, France}{bruno.guillon@uca.fr}{0000-0003-1630-3404}{}
\author{Luca Prigioniero}{Department of Computer Science, Loughborough University, Epinal Way, Loughborough LE11 3TU, UK}{l.prigioniero@lboro.ac.uk}{0000-0001-7163-4965}{}
\author{Javad Taheri}{Université Clermont Auvergne, Clermont Auvergne INP, LIMOS, CNRS, F\=/63000 Clermont-Ferrand, France}{javad.taheri369@gmail.com}{0009-0007-5106-2243}{}
\authorrunning{B. Guillon, L. Prigioniero, and J. Taheri}
\Copyright{Bruno Guillon, Luca Prigioniero, and Javad Taheri}
\date{April 2025}

\begin{document}
\maketitle
\begin{abstract}
	We investigate the descriptional complexity of different variants of $1$\=/limited automata (\las),
	an extension of two-way finite automata (\twnfas) characterizing regular languages.
	In particular,
	we consider \twnfas with common-guess (\twnfacgs),
	which are \twnfas equipped with a new kind of nondeterminism 
	that allows the device to initially annotate each input symbol, 
	before performing a read-only computation over the resulting annotated word.
	Their deterministic counterparts,
	namely two-way deterministic finite automata with common-guess (\twdfacgs),
	still have a nondeterministic annotation phase and
	can be considered as a restriction of \las.

	We prove exponential lower bounds for the simulations of \twdfacgs
	(and thus of \las)
	by deterministic \las 
	and by \twnfas.
	These results are derived from a doubly exponential lower bound for the simulation of \twdfacgs
	by one-way deterministic finite automata (\owdfas).
	Our lower bounds are witnessed by unary languages,
	namely languages defined over a singleton alphabet.
	As a consequence,
	we close a question left open in~[Pighizzini and Prigioniero. Limited automata and unary languages. Inf. Comput., 266:60-74],
	about the existence of a double exponential gap between \las and \owdfas in the unary case.
	Lastly, we prove an exponential lower bound
	for complementing unary \twdfacgs (and thus unary \las).
\end{abstract}

\section{Introduction}
The role of nondeterminism in computational models has been a
fundamental question in theoretical computer science since its inception.
In some computational models,
nondeterminism enhances computational power, for instance, in pushdown automata.
In others, it reduces the size of the description of the recognized language,
\eg, determinizing a finite automaton increases exponentially the size of the model,
and this blowup cannot be avoided in the worst case.
Understanding the impact of nondeterminism on the size of the description of computational models
is a central issue in descriptional complexity of formal systems.
The most famous problem in this area
is arguably the Sakoda and Sipser problem,
which conjectures an unavoidable super-polynomial blowup
for simulating \emph{two-way nondeterministic finite automata} (\twnfas)
with their deterministic counterparts (\twdfas)~\cite{SS78}.
Even the cost of the simulation of a classical (\ie, one-way) nondeterministic finite automata (\ownfas)
by \twdfas is unknown in the general case.%
\footnote{For completeness, we mention that this question has been solved in the unary case (see~\cite{Pig15}).}

A similar question arises in other models characterizing regular languages,
and in particular, when considering \emph{1\=/limited automata} (\las)
and their deterministic counterparts (\dlas).
These computational models extend \twnfas and \twdfas
by allowing them to rewrite the contents of the tape
under a severe restriction:
each tape cell can be rewritten
\emph{only when it is visited for the first time}.
It has been shown that providing two-way finite automata with this limited rewriting capability does not increase their expressive power:
$\la$s characterize regular languages~\cite{Wag89}.
As in the case of two-way finite automata,
the exact cost of the elimination of nondeterminism from \las remains unknown
--- see~\cite{Pig19} for a recent survey.
In~\cite{PP14},
the authors investigated the model from a descriptional complexity perspective.
They showed a tight exponential size cost
for the conversion of \dlas into one-way deterministic finite automata (\owdfas),
as well as a tight doubly-exponential size cost
for the conversion of \las into \owdfas.
Concerning the cost of determinizing \las (\ie, simulating \las with \dlas),
the just-exposed tight bounds imply a single exponential lower bound,
and an doubly-exponential upper bound.
Whether a single exponential would be sufficient in all case is still an open problem as of today, \cite[Problem~2]{Pig19}.

It is worth noting that the above-mentioned lower bounds obtained in~\cite{PP14}
are witnessed by non-unary languages,
\ie, languages over an alphabet having more than one symbol.
In many cases, the cost of simulations between computational models decreases
when considering the restricted case of unary languages (see, \eg,~\cite{Pig15}).
In~\cite{PP19},
working on this issue,
the authors showed
that the smallest \twnfa (and thus \owdfa) simulating a given $n$\=/state \dla
may require a number of states which is exponential in~$n$,
even when the input alphabet has a single letter.
They however left as an open question
whether the double-exponential lower bound for the conversion of \las into \owdfas
is still possible in the unary case~\cite[Problem~1]{Pig19}.

\bigskip

Often, nondeterminism plays a twofold role in \las:
in choosing the symbols that are written onto the tape
and in changing states when reading from it.
In~\cite{PP14},
the authors attributed the double exponential gap between \las and \owdfas
to this double role of nondeterminism in \las.
In order to investigate this phenomenon,
several restrictions of \las,
which somehow reduce the impact of nondeterminism in writing,
have been considered:
\emph{forgetting \las}~\cite{PP23a},
\emph{once-marking} and \emph{always-marking \las}~\cite{PP23b}.
We consider another
one\xspace
,
called \emph{two-way automata with common-guess} (\twnfacg and \twdfacg),
and we investigate it from a descriptional complexity perspective.
The model restricts \las by syntactically separating an initial write-only phase from a succeeding read-only phase.
More precisely, a \twnfacg proceeds in two consecutive modes over a given input~$w$:%
\begin{enumerate}[nosep]
	\item\strong{write-only annotation phase:}
		The machine nondeterministically annotates each letter of~$w$ in a memoryless write-only way using an auxiliary alphabet~$\Gamma$,
		known as the \emph{annotation alphabet}.
		This phase produces an annotated word~$x\in{(\Sigma\times\Gamma)}^*$ satisfying~$\pi_1(x)=w$,
		where~$\pi_1$ is the natural projection of~${(\Sigma\times\Gamma)}^*$ onto~${\Sigma}^*$.
	\item\strong{read-only validation phase:}
		The machine performs a read-only computation over the annotated word~$x$.
		For \twdfacg, this second phase (but not the preceding one) is required to be deterministic.
\end{enumerate}
Formally, a \twnfacg (\resp \twdfacg) $\machine{A}$ is defined as a triple $\langle\machine{M},\Sigma,\Gamma\rangle$
where~$\Sigma$ and~$\Gamma$ are respectively the \emph{input} and \emph{annotation alphabets},
and~$\machine{M}$ is a \twnfa (\resp a~\twdfa) over the product alphabet $\Sigma\times\Gamma$.
The machine~\machine{A} accepts a word $w\in\Sigma^*$
\ifof there exists an annotated word~$x\in{(\Sigma\times\Gamma)}^*$ accepted by~\machine{M} such that~$w=\pi_1(x)$.
Equivalently, $\langof{\machine{A}}=\pi_1(\langof{\machine{M}})$.
Remark that~\twdfacgs are nondeterministic devices,
since their annotation phase remains nondeterministic.

\emph{Common-guess} is a natural nondeterministic feature in other formalisms,
such as \emph{two-way transducers} (\ie, two-way automata that outputs words),
and \emph{MSO-transductions} (a formalism describing transformations using monadic second-order logic).
Indeed, in that context it has been shown that
the expressiveness of {two-way deterministic transducers with common guess}
coincides with that of {word-to-word nondeterministic MSO-transductions},
but is incomparable with that of {two-way nondeterministic transducers}~\cite{BDGP17}.
In fact, common-guess naturally corresponds to the nondeterminism of MSO-transductions,
since nondetermistic MSO-transductions are precisely deterministic MSO-transductions
operating over nondeterministically annotated (or coloured) inputs.
Concerning \twnfacgs and \twdfacgs,
the two models have already been considered in~\cite{GP19},
where simulations of \dlas by \twdfacgs, and of \las by \twnfacgs,
both with polynomial size cost,
have been shown.
Also, a doubly exponential lower bound for the simulation of \twdfacgs (and consequently of~\las) by \owdfas
(and hence an exponential lower bound for the simulation of \twdfacgs by \dlas or \twnfas)
has been obtained using a two-letter alphabet.
Hence, even when restricting the nondeterminism to the write role of \las,
the gap with \owdfas remains doubly exponential.

\subparagraph{Our results.}
In this paper we show that all the above-mentioned lower bounds for the various simulations of \las
still hold when the input alphabet is unary
and when furthermore the simulated \la is actually a \twdfacg
(see~\cref{tbl:lowerbounds}).
This includes a double-exponential lower bound for the conversion of unary \twdfacgs into \owdfas (\cref{thm:2dfa+cg for M_n,thm:1dfa for M_n} solving~\cite[Problem~1]{Pig19}),
from which an exponential lower bound for the conversion of unary \twdfacgs into \dlas is derived.
We also show an exponential lower bound in size for the transformation of unary \dlas into equivalent \twnfas.
Contrary to the analogous bound given on the number of states
in~\cite{PP19},
our bound relies on \dlas that use a constant number of work symbols only (\cref{thm:gap d1-la -> 2nfa}).
Finally, we show an exponential lower bound
for the transformation that converts a \twdfacg recognizing a unary language~$\lang{L}$
into a \la recognizing the complement of~$\lang{L}$ (\cref{thm:1-la for a*-M_n}).%
\begin{table}[t]%
	\centerline{\usetikzlibrary{matrix}
\begin{tikzpicture}
	\matrix[%
	matrix of nodes,
	row sep=-\pgflinewidth,
	column sep=-\pgflinewidth,
	nodes={text height=2ex, text depth=.5ex, text width=9.5em, align=center, draw=black, rectangle},
	column 1/.style={nodes={text width=5em}},
	row 1/.style={nodes={align=center}},
	] (m)	{
		{}				&	\owdfa														&	\twnfa															&	\dla														\\
		\dla			&	exp																&	exp~(\cref{thm:gap d1-la -> 2nfa})	&	-																\\
		\twdfacg	&	expexp~(\cref{thm:1dfa for M_n})	&	exp																	&	exp~(\cref{cor:d1-la for M_n})	\\
		\twnfacg	&	expexp														&	exp																	&	exp															\\
		\la				&	expexp														&	exp																	&	exp															\\
	};
	\path[every node/.style={sloped,inner sep=1.25pt}]
	(m-1-1.north west)	edge[draw=black]	(m-1-1.south east)
	(m-1-1.south west)	node[above right] {source}
	(m-1-1.north east)	node[below left]	{target}
	;
	\path[
		transform shape,
		every edge/.style={draw=none,fill=none},
		imply back/.style={fill=white},
		imply back left/.style={imply back,minimum height=1\baselineskip},
		imply back down/.style={imply back,minimum width=4\baselineskip},
		imply/.style={fill=none,inner sep=0pt,node contents={$\implies$},anchor=mid},
		imply left/.style={imply,rotate=180},
		imply down/.style={imply,rotate=270,xscale=.5,yscale=1.5},
	]
	(m-2-3)	edge	node[imply back left]{}	node[imply left]{} (m-2-2)	
	(m-3-2)	edge	node[imply back down]{}	node[imply down]{}	(m-4-2)
	(m-4-2)	edge	node[imply back down]{}	node[imply down]{}	(m-5-2)
	(m-2-3)	edge	node[imply back down]{}	node[imply down]{}	(m-3-3)
	(m-3-3)	edge	node[imply back down]{}	node[imply down]{}	(m-4-3)
	(m-4-3)	edge	node[imply back down]{}	node[imply down]{}	(m-5-3)
	(m-3-4)	edge	node[imply back down]{}	node[imply down]{}	(m-4-4)
	(m-4-4)	edge	node[imply back down]{}	node[imply down]{}	(m-5-4)
	;
\end{tikzpicture}}%
	\caption{Lower bounds for the simulations between variants of \las in the unary case.
	The simulations are from the source machines
	listed in rows,
	to the target devices
	listed in columns.
	Each cell contains the size lower bound for the simulation,
	with an indication of where the bound is derived from (ingoing arrow)
	or the result in this paper where it is proved.}
	\label{tbl:lowerbounds}
\end{table}

\subparagraph{Outline.}
The paper is organized as follows.
In \cref{sec:prelim} we present the main definitions and the basic properties of the considered computational models,
which are used in the subsequent sections.
In \cref{sec:binseq} we introduce a family~${(\fbs)}_{n\in\Nat}$ of singleton languages 
defined over a three-letter alphabet,
which plays a central role in our results.
Using this family of languages,
we show that unary \twdfacgs and unary \dlas are exponentially more succinct than \twnfas in \cref{sec:d1-la to 2nfa}.
Then, building on top of~\fbs
in \cref{sec:2dfa+cg to d1-la},
we define a family of infinite unary languages~${(\lang{M}_n)}_{n\in\Nat}$,
which allows us to derive the lower bounds
for the simulations of \twdfacgs (and consequently of \las) by \owdfas, \ownfas, and \dlas.
Finally, in \cref{sec:complement}, we study the complement of~$\lang{M}_n$
and show that it requires a double exponential size in~$n$ to be recognized by a \ownfa
and thus an exponential size in~$n$ to be recognized by a \la.

\section{Preliminaries}
\label{sec:prelim}
In this section, we recall some fundamental definitions and notations used throughout the paper.
We assume that the reader is familiar with basic concepts from formal languages and automata theory
(see, \eg,~\cite{HU79}).
For a set $S$, $\defem{\card{S}}$ denotes its cardinality.
Given an alphabet $\Sigma$,
the set of strings over~$\Sigma$
is denoted by~\defem{$\Sigma^*$}.
It includes the empty string denoted by~\defem{\ew}.
The length of a word~$w\in\Sigma^*$ is denoted by~\defem{$\lvert w\rvert$}.
For a language~$\lang{L}\subseteq\Sigma^*$,
we denote by~\defem{$\pref{\lang{L}}$}, \defem{$\suff{\lang{L}}$}, and~\defem{$\factor{\lang{L}}$}
the languages of prefixes, suffixes, and factors of (words of)~$\lang{L}$,
respectively.

\begin{definition}
	\label{def:2nfa}
	A \defem{two-way nondeterministic finite automaton} (\defem{\twnfa}) is a tuple
	$\machine{A}=\braket{Q,\Sigma,\delta,q_0,F}$,
	where~$Q$ is the finite set of states,
	$\Sigma$ is the input alphabet,
	$q_0\in Q$ is the initial state,
	$F\subseteq Q$ is the set of final states,
	and~$\delta:Q\times\Sigma_{\lend\rend}\rightarrow 2^{Q\times\set{-1,+1}}$
	is a nondeterministic transition function
	with ${\Sigma_{\lend\rend}=\Sigma\cup\set{\lend,\rend}}$,
	where $\lend,\rend\notin\Sigma$
	are two special symbols called \defem{the left} and \defem{the right endmarker}, respectively.
\end{definition}
In \twnfas,
the input is written on the tape surrounded by the two endmarkers,
the left endmarker being at position zero.
Hence, on input~$w$, the right endmarker is at position~${\lvert w \rvert+1}$.
In one move,
\machine{A} reads an input symbol,
changes its state,
and moves the head one position backward or forward depending on whether~$\delta$ returns~$-1$ or~$+1$, respectively.
Furthermore,
the head cannot pass the endmarkers.
The machine accepts the input
if there exists a computation path starting from the initial state~$q_0$
with the head on the cell at position~$1$
(\ie, scanning the first letter of~$w$ if~$w\neq\emptyword$ and scanning~$\rend$ otherwise)
and eventually halts in a final state~$q_f\in F$ with the head scanning the right endmarker.
The language accepted by \machine{A} is denoted by $\langof{\machine{A}}$.

A \twnfa is said to be \defem{deterministic} (\defem{\twdfa})
whenever $\card{\delta(q,\sigma)}\leq 1$,
for every $q\in Q$ and $\sigma\in\Sigma\cup\set{\lend,\rend}$.
It is called \defem{one-way} if its head can never move backward, \ie, if no transition returns $-1$.
By \defem{\ownfa} and \defem{\owdfa} we denote one-way nondeterministic and deterministic finite automata, respectively.

The above models are all read-only machines.
We now extend \twnfas with a limited write ability.%
\begin{definition}
	\label{def:1-la}
	A \defem{$1$-limited automaton} (\la, for short)
	is a tuple $\machine{A}=(Q,\Sigma,\Delta,\delta,q_0,F)$,
	where~$Q$, $\Sigma$, $q_0$, and~$F$ are defined as for~\twnfas,
	the finite set~$\Delta$ is the \defem{work alphabet} which includes the input alphabet~$\Sigma$ but excludes~$\set{\lend,\rend}$,
	and the nondeterministic transition function is~%
	$\delta:Q\times\Delta_{\lend\rend}\rightarrow 2^{Q\times\Delta_{\lend\rend}\times\set{-1,+1}}$
	with~$\Delta_{\lend\rend}=\Delta\cup\set{\lend,\rend}$.
\end{definition}
In one move, according to~$\delta$,
\machine{A} reads a symbol from the tape,
changes its state, replaces the symbol just read by a new symbol from the work alphabet,
and moves its head one position backward or forward.
However, replacing symbols is subject to some restrictions, which, essentially,
allow to modify the content of a cell during the first visit only.
Technically, symbols from~$\Sigma$ shall be replaced with symbols from~$\Delta\setminus\Sigma$,
while symbols from~$\Delta_{\lend\rend}\setminus\Sigma$ are never changed.
In particular, at any time, both special symbols~\lend and~\rend
occur exactly once on the tape and exactly at the respective left and right boundaries.
We say that~$\machine{A}$ is \defem{deterministic} (\defem{\dla})
if $\card{\delta(p,\gamma)}\leq1$ for each~$(p,\gamma)\in Q\times\Delta_{\lend\rend}$.
Acceptance for a~\las is defined exactly as for \twnfas:
the machine accepts an input~$w$ if, starting in state~$q_0$ with the head scanning the position~$1$ of the tape whose contents is~$\lend w\rend$,
it eventually enters a final state~$q\in F$ with the head scanning the right endmarker.
The language accepted by a given \la~$\machine{A}$ is denoted by~$\langof{\machine{A}}$.

\subparagraph{Common-guess.}
We now extend \twnfas and \twdfas with \emph{common-guess}.
This feature describes the ability to initially annotate the input word~$w\in\Sigma^*$
using some annotation symbols from a fixed alphabet~$\Gamma$,
to which we refer as the \defem{annotation alphabet}.
The annotated word resulting from this initial phase
is a word over the product alphabet~$\Sigma\times\Gamma$.
It is nondeterministically chosen among all the words~$v\in{(\Sigma\times\Gamma)}^*$
such that~$\pi_1(v)=w$ (in particular~$\lvert v \rvert =\lvert w\rvert$),
where~\defem{$\pi_1$} is the natural projection of~${(\Sigma\times\Gamma)}^*$ onto~$\Sigma^*$.%
\begin{definition}
	\label{def:2nfa+cg}
	A {\twnfa} (\resp \twdfa) \emph{with common-guess} (\defem{\twnfacg}, \resp{} \defem{\twdfacg})
	is a triplet ${\machine{M}=\braket{\machine{A},\Sigma,\Gamma}}$
	where~$\Sigma$ is the input alphabet,
	$\Gamma$ is the \defem{annotation alphabet},
	and~\machine{A} is a \twnfa (\resp \twdfa) over the product alphabet~$\Sigma\times\Gamma$.
\end{definition}
The language recognized by~$\machine{M}$ is:
\[
	\langof{\machine{M}}=\set{w\in\Sigma^*\mid \exists v\in{(\Sigma\times\Gamma)}^*,\: v\in\langof{\machine{A}}\mathinner{\text{ and }}\pi_1(v)=w}=\pi_1(\langof{\machine{A}})
	\text.
\]
It is worth noting that,
because the annotation phase is nondeterministic by nature (witnessed by the $\exists$~quantifier above),
\twdfacgs are nondeterministic devices.

\subparagraph{Size of models.}
For each model under consideration,
we evaluate its size as the total number of symbols
used to describe it.
Hence, under standard representation and denoting by~$\Sigma$ the input alphabet,
the \defem{size} of an $n$\=/state \twnfa is~$\bigO(n^2\card{\Sigma})$,
that of an $n$\=/state \la with work alphabet~$\Delta$ is~$\bigO(n^2\card{\Delta}^2)$,
and that of an $n$\=/state \twnfacg with annotation alphabet~$\Gamma$ is~$\bigO(n^2\card{\Sigma} \cdot \card{\Gamma})$.
In our work, we generally consider~$\card{\Sigma}$ as a constant (which is often equal to~$1$).

\subparagraph{\twnfacgs underlying \twnfas.}
By definition, in order to define a \twnfacg over~$\Sigma$,
it suffices to describe its underlying \twnfa over~$\Sigma\times\Gamma$.
Yet,
when~$\Sigma=\set{a}$ is unary,
$\set{a}\times\Gamma$ is isomorphic to~$\Gamma$.
Hence, we will not distinguish the triple~$\braket{\machine{A},\set{a},\Gamma}$
in which~$\machine{A}$ is a \twnfa over~$\Gamma$
from the \twnfacg~$\braket{\machine{A'},\set{a},\Gamma}$
in which~$\machine{A'}$ is the \twnfa isomorphic to~$\machine{A}$,
obtained by replacing each symbol~$\gamma\in\Gamma$ by~$(a,\gamma)$
in the transitions of~$\machine{A}$.
Given a language~$\lang{L}$,
we call \emph{language of lengths of~$\lang{L}$}
the language:
\[
	\defem{\norm{\lang{L}}}=\Set{a^\ell \mid \exists u\in\lang{L},\:{\lvert u\rvert}=\ell}
\]
The above comment directly gives the following result.%
\begin{proposition}
	\label{prop:2nfa+cg for norm}
	If a language~$\lang{L}\subseteq\Gamma^*$ is recognized by a~\twnfa~$\machine{A}$,
	then the unary language $\norm{\lang{L}}$
	is recognized by a~\twnfacg~$\machine{B}$
	with the same state set as~$\machine{A}$
	and annotation alphabet~$\Gamma$.
	Furthermore, if~$\machine{A}$ is deterministic, then~$\machine{B}$ is a \twdfacg.
\end{proposition}

\subparagraph{Relating \twnfacgs and \twdfacgs to \las.}
We now briefly discuss how \twnfacgs and \twdfacgs are related to \las.
It is routine to simulate an $n$\=/state \twnfacg~$\braket{\machine{A},\Sigma,\Gamma}$
with an $(n+1)$\=/state\xspace%
\footnote{%
	The additional state,
	which is the initial state,
	is used to annotate the input tape at the beginning.
	Since after the implied left-to-right traversal of the tape,
	the tape symbols are no longer input symbols from~$\Sigma$
	but annotated symbols from~$\Sigma\times\Gamma$,
	the same state can be used to move the head back to the left endmarker
	and then start the read-only simulation of the \twdfa underlying the simulated \twdfacg.%
	%
}
\la with work alphabet~$\Sigma\times\Gamma$.
Hence~\twnfacgs and~\twdfacgs can be seen as particular cases of \las.
In~\cite{GP19}, the authors showed that a polynomial increase in size
is always sufficient for the conversion of \la into \twnfacg,
as well as the conversion of \dla into~\twdfacg.
However, an exponential gap was observed for the conversion of \twdfacg into \dla.
Yet, in some particular cases,
we can turn a \twdfacg into a \dla of similar size.
For this to be possible,
a sufficient condition is that,
when entering a cell for the first time,
at most one annotation symbol~$\gamma\in\Gamma$ for that cell allows the machine to continue its computation.
Indeed, in such a case,
instead of reading the annotated symbol from the tape,
the simulating \dla reads the input symbol~$\sigma$ and rewrites it with~$(\sigma,\gamma)$
where~$\gamma$ is the unique annotation symbols allowing the computation to proceed.
This property is formalized by the concept of \emph{$\Gamma$\=/predictive}~ \twdfas.%
\begin{definition}
	\label{def:predictive}
	Let~$\machine{A}=\braket{Q,\Sigma\times\Gamma,\delta,q_0,F}$ be \twdfa over~$\Sigma\times\Gamma$.
	We say that~$p\in Q$ is \defem{$\Gamma$\=/predictive}
	if for every~$\sigma\in\Sigma$,
	there exists at most one~$\gamma\in\Gamma$
	such that~$\delta(p,(\sigma,\gamma))\neq\emptyset$.
	The \twdfa~$\machine{A}$ is said~\defem{$\Gamma$\=/predictive},
	if for every word~$w\in\langof{\machine{A}}$,
	each time the machine enters a cell for the first time during its computation over~$w$,
	the current state is $\Gamma$\=/predictive.
\end{definition}

As announced, $\Gamma$\=/predictive \twdfacgs over~$\Sigma$
can be turned into equivalent \dlas
with same number of states and work alphabet~$\Sigma\times\Gamma$.%
\begin{proposition}
	\label{prop:2dfa to d1-la}
	If \machine{A} is a $\Gamma$\=/predictive \twdfa over $\Sigma\times\Gamma$,
	then there exists a \dla over~$\Sigma$ with the same state set as \machine{A} and work alphabet $\Sigma\cup(\Sigma\times\Gamma)$,
	which recognizes $\pi_1(\langof{\machine{A}})$.
\end{proposition}%
\begin{proof}[Proof sketch]
	Let~$\machine{A}=\braket{Q,\Sigma\times\Gamma,\delta,q_0,F}$,
	$\Delta=\Sigma\cup(\Sigma\times\Gamma)$,
	and $\Delta_{\lend\rend}=\Delta\cup\set{\lend,\rend}$.
	We define~$\delta'$ from~$Q\times\Delta_{\lend\rend}$ to subsets of~$Q\times(\Delta_{\lend\rend}\setminus\Sigma)\times\set{-1,+1}$
	as follows.
	For each~$p\in Q$ and~$\tau\in\Delta_{\lend\rend}$, $\delta'(p,\tau)$ is equal:
	\begin{itemize}[nosep]
		\item to~$\set{(q,\tau,d)|(q,d)\in\delta(p,\tau)}$ if~$\tau\notin\Sigma$;
		\item to~$\set{(q,(\tau,\gamma),d)|\gamma\in\Gamma,\;(q,d)\in\delta(p,(\tau,\gamma)}$ if~$\tau\in\Sigma$ and~$p$ is $\Gamma$\=/predictive;
		\item to~$\emptyset$ otherwise.
	\end{itemize}
	By construction and determinism of~$\machine{A}$, $\card{\delta'(p,\tau)}\leq1$ for each~$p\in Q$ and~$\tau\in\Delta_{\lend\rend}$.
	We let~$\machine{B}$ be the \dla~$\braket{Q,\Sigma,\Delta,\delta',q_0,F}$.
	It is routine to prove that~$\machine{B}$ is equivalent to~$\machine{A}$,
	using the fact that~$\machine{A}$ is~$\Gamma$\=/predictive.
\end{proof}

As before,
we sometimes do not distinguish a \twdfa over~$\Gamma$ from a \twdfa over~$\set{a}\times\Gamma$
since the two alphabets are isomorphic.
We may thus say that a \twdfa over~$\Gamma$ is $\Gamma$\=/predictive.
This allows us to transform it into a unary \dla, according to \cref{prop:2dfa to d1-la}.

\section{Full binary sequence}
\label{sec:binseq}
Our main results establish exponential and double-exponential lower bounds
for various simulations of \las, \twdfacgs, and \dlas by two-way and one-way finite automata.
All these results are obtained over a unary alphabet.
Yet, as the simulated devices are allowed to rewrite their tape contents,
the witness languages actually rely on a particular word~$\bin*$ over a three-letter alphabet,
called \emph{full binary sequence}.
The word~$\bin*$ has length longer than~$2^n$.
Indeed, it consists of consecutive encodings of integers in base~$2$ over~$n$ bits,
separated by a special marker~$\ssep$.
Formally, for a fixed positive integer~$n$ and for each~$i=0,\ldots,2^n-1$,
we denote by~\defem{$\bin{i}$} the binary representation of~$i$ over~$n$ bits,
with the most significant bit on the left.
Hence, $\set{\bin{i}\mid0\leq i<2^n}=\set{0,1}^n$.
We define the \defem{full binary sequence} as follows:%
\begin{align*}
	\defem{\bin*} &= \bin0 \ssep \bin1 \ssep \cdots \ssep \bin{2^n-1} \ssep
\end{align*}
Its length is~$2^n(n+1)$.
For instance, $b_3=000\ssep001\ssep010\ssep011\ssep100\ssep101\ssep110\ssep111\ssep$.
Let us define the language~$\defem{\fbs}=\set{\bin*}$ containing only the word~$\bin*$.
We shall show that this language can be recognized by a \twdfa with a number of states linear in~$n$.
This will be obtained using the following observation.%
\begin{fact}
	\label{fact:successor bin}
	For every positive integer~$n$ and every $0\leq i,j\leq 2^n-1$,
	it holds that~$j=i+1$ \ifof
	there exist an integer~$m\in\mathbb{N}$ and a word~$x\in\set{0,1}^*$
	such that~$\bin{i}=x01^m$ and~$\bin{j}=x10^m$.
\end{fact}

As a direct consequence of \cref{fact:successor bin}
one can notice that the last symbol of each length-$(n+2)$ factor of~$\bin*$
is uniquely determined by its~$n+1$ preceding symbols.
Moreover, it is easy to see that the factor~$1^n\ssep$ is the length-$(n+1)$ suffix of~$\bin*$.
This is formalized in the following.
For each symbol~$\sigma\in\set{0,1,\ssep}$,
we define the regular language~$X_\sigma$ describing the pattern matching the~$n+1$ symbols preceding~$\sigma$ in a factor of~$\bin*$,
and we define the regular language~$X_\dsep$ describing the pattern matching the length-$(n+1)$ suffix of~$\bin*$:%
\begin{align*}
	\begin{array}{rclll}
		\defem{X_0}				&\!\!\!\!=\!\!\!\!&	1{1}^*\ssep{(0+1)}^+ &\!\!+\!\!& 0{1}^*0{(0+1)}^*\ssep{(0+1)}^*\text, \\
		\defem{X_1}				&\!\!\!\!=\!\!\!\!&	0{1}^*\ssep{(0+1)}^* &\!\!+\!\!& 1{1}^*0{(0+1)}^*\ssep{(0+1)}^*\text,
	\end{array}
	&&
	\begin{array}{rcl}
		\defem{X_\ssep}	&\!\!\!\!=\!\!\!\!&	\ssep{(0+1)}^+\text,
		\\
		\defem{X_\dollar}	&\!\!\!\!=\!\!\!\!&	1^+\ssep
		\text.
	\end{array}
\end{align*}
These four languages are disjoint.
The membership of a length-$(n+1)$ factor~$u$ of~$\bin*$ to some~$X_\sigma$, for~$\sigma\in\set{0,1,\ssep,\dsep}$,
determines the unique symbol~$\sigma$ that can be appended to~$u$ to form a factor of~$\bin*\dsep$
(where~$\dsep$ is used to detect the end of the word).%
\begin{lemma}
	\label{lem:unique extension of factors of b_n}
	Let~$\Sigma=\set{0,1,\ssep}$,
	and let~$X_0$, $X_1$, $X_\ssep$, and~$X_\dsep$
	be the four disjoint regular languages defined above.
	For every positive integer~$n$,
	every length-$(n+1)$ word~$u\in\Sigma^*$,
	and every~$\sigma\in\Sigma\cup\set{\dsep}$,
	$u\sigma$ is a factor of~$\bin*\dollar$ \ifof $u\in X_\sigma$.
\end{lemma}

Notably, these languages do not depend on~$n$.
Indeed, it can be noticed that
their union~$X_0\cup X_1\cup X_\ssep\cup X_\dollar$
includes all length-$(n+1)$ factors of~$\bin*$ for every positive integer~$n$.
Conversely, every word~$w$ belonging to this union
has length at least~$2$ and is a factor of~$\bin*$,
where~$n=\lvert w\rvert-1$.
Also, one of the consequences of \cref{lem:unique extension of factors of b_n}
is that every length-$(n+1)$ factor~$u$ of~$\bin*$
occurs exactly once in~$\bin*$,
and thus identifies the prefix and the suffix of~$\bin*$ that ends or starts with~$u$.%
\begin{proposition}
	\label{prop:factor identify prefix}
	\label{prop:factor identify suffix}
	\label{prop:factors identify factor}
	Let~$n>0$, $\Sigma=\set{0,1,\ssep}$, $x,y\in\Sigma^*$, and~$u,v\in\Sigma^{n+1}$.
	\begin{enumerate}[nosep]
		\item\label{it:factor identify suffix}If~$ux$ and~$uy$ are suffixes of~$\bin*$, then~$x=y$.
		\item\label{it:factor identify prefix}If~$xv$ and~$yv$ are prefixes of~$\bin*$, then~$x=y$.
		\item\label{it:factors identify factor}If~$x$ and~$y$ are factors of~$\bin*$
			which both admit~$u$ as prefix and~$v$ as suffix
			(possibly with overlapping),
			then~$x=y$.
	\end{enumerate}
\end{proposition}%
\begin{proof}
	We first prove \cref{it:factor identify suffix} by induction on~$\lvert x\rvert$,
	assuming, \Wlog[w], $\lvert x\rvert \geq \lvert y\rvert$.
	If~$\lvert x\rvert=0$ then both~$x$ and~$y$ are empty and thus equal.
	Otherwise, $x$ is non-empty.
	Let~$\sigma$ be the first symbol of~$x$.
	Since~$ux$ is a suffix of~$\bin*$, $u\sigma$ is a factor of~$\bin*$,
	and thus, by \cref{lem:unique extension of factors of b_n}, $u\in X_\sigma$.
	Therefore, $y$ also starts with~$\sigma$.
	We conclude by induction using the length-$(n+1)$ factor~$u'\sigma$,
	where~$u'$ is the suffix of length~$n$ of~$u$.

	Now, for proving \cref{it:factor identify prefix},
	we suppose that~$xu$ and~$yu$ are prefixes of~$\bin*$.
	Thus, there exist~$x'$ and~$y'$ such that~$\bin*=xux'=yuy'$.
	By the proof above, $x'=y'$ and thus~$x=y$.

	Finally, for proving \cref{it:factors identify factor},
	we suppose that~$x$ and~$y$ are factors of~$\bin*$.
	Therefore, there exist~$x'$, $x''$, $y'$, and~$y''$
	such that~$\bin*=x'xx''=y'yy''$.
	By \cref{it:factor identify prefix,it:factor identify suffix},
	$x'=y'$ (because~$xx''$ and~$yy''$ are both suffixes starting with~$u$ and thus equal)
	and~$x''=y''$ (because~$x'x$ and~$y'y$ are both prefixes ending with~$v$ and thus equal).
	Hence~$x=y$.
\end{proof}

As seen before, $X_0$, $X_1$, $X_\ssep$ and~$X_\dsep$ are disjoint
and contain only words of length larger than~$1$ over~$\Sigma=\set{0,1,\ssep}$.
We define~\defem{$X_\bot$} to be the regular language of words over~$\Sigma$
that are not prefixes of any word in~$X_0\cup X_1\cup X_\ssep\cup X_\dsep$.
Each of~$X_0$, $X_1$, $X_\ssep$, $X_\dsep$, and~$X_\bot$
is a regular language that does not depend on~$n$,
and hence is recognized by a \owdfa of constant size.
By taking the product of these \owdfa\s,
we obtain a \owdfa able to recognize each of the above languages
(using a specific accepting state~$q_\sigma$ for each~$\sigma\in\Sigma\cup\set{\dsep,\bot}$).
Such an automaton, using~$11$ states only, is given in \cref{fig:1DFA:product X}.
We call it~\defem{$F$}.%
\begin{figure}[thb]
	\Centering
	\includegraphics[page=1]{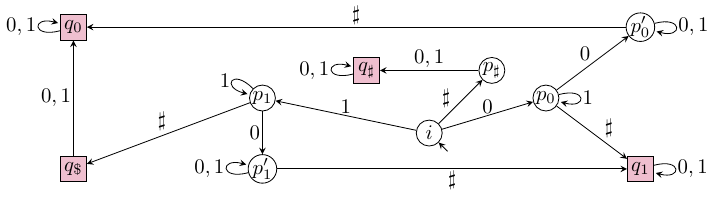}
	\caption{%
		The $11$\=/state automaton~$F$ recognizing~$X_\sigma$,
		for each~$\sigma\in\set{0,1,\ssep,\dollar,\bot}$
		using the corresponding state~$q_\sigma$
		(represented as a purple-filled rectangle);
		the trap state~$q_\bot$ (entered from every state~$q_\sigma$ or~$p_\ssep$ by reading~$\ssep$) is not depicted.%
	}
	\label{fig:1DFA:product X}
\end{figure}

In order to force the length of the recognized factor to be~$n+1$,
we take the product of~$F$ with the canonical $(n+2)$-state \owdfa recognizing~$\Sigma^{n+1}$.
This leads to a $11(n+2)$-state \owdfa~\defem{$F_n$}
that determines, using distinguished final states,
whether a given input has length~$n+1$ and belongs to~$X_0$, $X_1$, $X_\ssep$, $X_\dsep$, or~$X_\bot$.%
\begin{lemma}
	\label{lem:1DFA for consecutive}
	Let~$n>0$ and~$\Sigma=\set{0,1,\ssep}$.
	There exists a $11(n+2)$-state \owdfa~$F_n$
	that has five distinguished states~$r_0,r_1,r_\ssep,r_\dsep,r_\bot$,
	such that,
	for each~$u\in\Sigma^*$ and each~${\sigma\in\Sigma\cup\set{\dsep,\bot}}$,
	$F_n$ goes from the initial state to~$r_\sigma$ on $u$
	\ifof $\lvert u\rvert=n+1$ and~$u\in X_\sigma$.
\end{lemma}
As we are interested only in factors with length~$n+1$,
we may assume that the five distinguished states~$r_0$, $r_1$, $r_\ssep$, $r_\dsep$, and~$r_\bot$
have no outgoing transition,
\ie, they are halting states.
We denote by~\consecutive{} the procedure
that simulates~$F_n$ until either reaching the end of the input, which is marked by~$\rend$,
or entering one of the states~$r_\sigma$ from~$\set{r_0,r_1,r_\ssep,r_\dollar,r_\bot}$.
In the former case, the procedure returns~$\rend$,
while in the latter case it returns~$\sigma$.
This allows us to design a \twdfa recognizing factors of~$\bin*$.%
\begin{procedure}[t]
	\caption{()\space
		\isFactBinSeq{}%
		\newline%
		Check that the input is a factor of the full binary sequence~$\bin*$.
		On success, the procedure halts on the right endmarker.%
	}
	\label{proc:isFactBinSeq}%
	\While{$\read()\neq\rend$}{%
		$\sigma\gets\consecutive{}$\tcp*[r]{the head is moved~$n+1$ cells to the right}
		\label{l:proc/isFactBinSeq:call consecutive}
		\lIf{$\sigma=\bot$}{\reject}
		\lIf{$\sigma=\rend$}{\label{l:proc/isFactBinSeq:incomplete sequence}\accept}%
		\lIf{$\sigma=\dollar$ \KwAnd $\read()=\rend$}{\label{l:proc/isFactBinSeq:complete sequence}\accept}%
		\lIf{$\sigma\neq\read()$}{\reject}
		\label{l:proc/isFactBinSeq:check sigma}
		move the head~$n$ cells to the left
		\label{l:proc/isFactBinSeq:left moves}
	}
\end{procedure}%
\begin{lemma}
	\label{lem:2dfa for Fact(b_n)}
	There exists a function~$f\in\bigO(n)$
	such that,
	for every~$n>0$,
	there exists a~$f(n)$-state \twdfa over~$\set{0,1,\ssep}$
	that recognizes~$\factor{\bin*}$,
	namely the language of factors of~$\bin*$.
\end{lemma}%
\begin{proof}
	The \twdfa implements \isFactBinSeq{} given in \cref{proc:isFactBinSeq}.
	A number of states linear in~$n$ is sufficient
	to implement \consecutive{}
	(called at \cref{l:proc/isFactBinSeq:call consecutive} and that simulates~$F_n$)
	according to \cref{lem:1DFA for consecutive},
	and to perform the left moves at \cref{l:proc/isFactBinSeq:left moves}.
	These states are local to these two lines,
	and hence independent.
	Furthermore, the variable~$\sigma$ ranges over the set~$\set{0,1,\ssep,\dollar,\bot,\rend}$ of constant size.
	Therefore,
	the \twdfa uses a number of states in~$\mathcal{O}(n)$.
\end{proof}

The \twdfa given by the lemma can then be adapted to recognize particular factors of~$\bin*$.
In particular, if we fix the~$n+1$ first symbols of the factors to recognize,
the \twdfa can be made $\Gamma$\=/predictive,
where~$\Gamma=\set{0,1,\ssep}$ is the input alphabet,
that is, each time an input cell is visited for the first time,
all but at most one symbol causes an immediate rejection of the input.%
\begin{lemma}
	\label{lem:2dfa for b_n}
	\label{lem:2dfa for b_n and special factors}
	Let~$\Gamma=\set{0,1,\ssep}$, $n\geq0$, and~$\ell\in\set{0,\ldots,2^n(n+1)}$.
	Then, there exist
	\begin{itemize}[nosep]
		\item a $\Gamma$\=/predictive \twdfa with $\bigO(n)$ states recognizing $\pref{\bin*}\cap\Gamma^\ell$;
		\item a $\Gamma$\=/predictive \twdfa with $\bigO(n)$ states recognizing $\suff{\bin*}\cap\Gamma^\ell$;
		\item a $\Gamma$\=/predictive \twdfa with $\bigO(n)$ states recognizing $\set{w\in\pref{\bin*} \mid \lvert w\rvert\geq \ell}$;
		\item a \twdfa with $\bigO(n)$ states recognizing $\set{w\in\suff{\bin*}\mid \lvert w\rvert \geq \ell}$.
	\end{itemize}
	In particular, $\set{\bin*}$ is recognized by a $\Gamma$\=/predictive \twdfa with~$\bigO(n)$ states.
\end{lemma}%
\begin{proof}
	If~$\ell<n+1$ the results are trivial.
	We thus suppose~$\ell\geq n+1$
	and we let~$u$ be a factor of~$\bin*$ of length~$\ell$,
	$x$ be the prefix of~$u$ of length~$n+1$,
	and~$y$ be the suffix of~$u$ of length~$n+1$
	($x$ and~$y$ might overlap).
	Our construction relies on the fact that the pair~$(x,y)$ uniquely identifies~$u$,
	according to \cref{it:factors identify factor} of \cref{prop:factor identify suffix}.

	We first build a $\Gamma$\=/predictive \twdfa~$\machine{A}_{x,y}$ with~$\bigO(n)$ states that recognizes~$\set{u}$.
	We let it operate in three successive stages:
	\smallskip
	\begin{enumerate}[nosep,label={Stage~\arabic*:},ref={Stage~\arabic*},labelwidth=3em,leftmargin=5em,font=\strongsty]
		\item\label{stage:check prefix} it checks that the input starts with~$x$;
		\item\label{stage:call isFactBinSeq} it simulates \cref{proc:isFactBinSeq} (from the initial position, namely position~$1$);
		\item\label{stage:check suffix} it checks that the input ends with~$y$.
	\end{enumerate}
	\smallskip
	Clearly, since both~$x$ and~$y$ are fixed words of length~$n+1$ and according to \cref{lem:2dfa for Fact(b_n)},
	a linear number of states in~$n$ is sufficient for our \twdfa~$\machine{A}_{x,y}$ to implement these three stages.
	We now argue that~$\machine{A}_{x,y}$ is~$\Gamma$\=/predictive.
	During \ref{stage:check prefix}, $\machine{A}_{x,y}$ visits only the ${n+1}$ first input cells.
	For each of them,  $\machine{A}_{x,y}$ compares the symbol it contains with the corresponding symbol in~$x$,
	and immediately rejects the input whenever the two symbols differ.
	Hence, for each~$i\leq n+1$, when visiting the cell at position~$i$ for the first time,
	only the~$i$\=/th symbol of~$x$ allows~$\machine{A}_{x,y}$ to proceed.
	After \ref{stage:check prefix},
	the machine moves the head back to the cell at position~$1$ and starts \ref{stage:call isFactBinSeq}.
	During this second stage, every input cell is visited,
	according to \cref{proc:isFactBinSeq}.
	For each~$i>n+1$, when the cell at position~$i$ is visited for the first time,
	it is entered during the last step of the call to \consecutive (\cref{l:proc/isFactBinSeq:call consecutive}),
	at which point the expected symbol~$\sigma$ is known (see \cref{l:proc/isFactBinSeq:check sigma}).
	Hence~$\machine{A}_{x,y}$ is~$\Gamma$\=/predictive.

	By forcing~$x=0^n\ssep$ (\resp $y=1^n\ssep$) in~$u$
	we obtain a $\Gamma$\=/predictive \twdfa with~$\bigO(n)$ states
	recognizing the language $\pref{\bin*}\cap\Gamma^\ell$ (\resp $\suff{\bin*}\cap\Gamma^\ell$).

	In order to recognize~$\set{w\in\pref{\bin*}\mid\lvert w\rvert\geq\ell}$
	it suffices to change \ref{stage:check suffix}
	so that, instead of checking that~$y$ is a suffix of the input,
	it checks that~$y$ appears as a factor.
	This can still be done with a linear number of states in~$n$,
	preserving the property of being $\Gamma$\=/predictive.

	In order to recognize~$\set{w\in\suff{\bin*}\mid\lvert w\rvert\geq\ell}$
	we proceed similarly,
	modifying \ref{stage:check prefix}
	so that, instead of checking that~$x$ is a prefix of the input,
	it checks that~$x$ appears as a factor.
	This can also be implemented with a linear number of states in~$n$,
	but the resulting automaton is not $\Gamma$\=/predictive.

	Finally, we observe that, by choosing~$\ell=2^n(n+1)$,
	we obtain $\set{\bin*}=\pref{\bin*}\cap\Gamma^\ell$.
\end{proof}

\section{Exponential gap from \dlas and \twdfacgs to \twnfas}
\label{sec:d1-la to 2nfa}
In this section, we show an exponential gap in size for the simulation of \dlas by~\twnfas.
This gap holds even in the particular case of unary languages.
Indeed, our witness language is~$\fbs=\set{\bin*}$.
A similar gap has already been proven for the simulation in~\cite{PP19},
where a family~${(\lang{L}_n)}_{n\in\Nat}$ of witness languages was presented
such that each~$\lang{L}_n$ is recognized by a~\dla~$\machine{M}_n$ having a number of states linear in~$n$,
but is not recognized by any \twnfa with less than~$2^n$ states.
However, the machine~$\machine{M}_n$ uses a work alphabet of size linear in~$n$,
making the total size of the machine quadratic in~$n$.%
\footnote{%
	As~$\machine{M}_n$ is a \dla,
	it is standard to describe it
	using~$n(\card{\Sigma}\!\cdot\!\card{\Gamma}+\card{\Gamma})\in\bigO(n^2)$ symbols.%
}
Therefore, though the observed gap in terms of state complexity is exponential,
it is sub-exponential if we consider the size of the machine, taking into account the work alphabet size.
Our construction --- which is much more naive and less elegant than those given in~\cite{PP19} ---
uses a work alphabet of constant size combined with a linear number of states in~$n$,
and thus allows us to derive an exponential gap in size.

We also prove that a linear number of states and a constant number of annotation symbols
are sufficient for \twdfacgs to recognize our witness languages.

\begin{lemma}
	\label{lem:2dfa+cg for b_n}
	\label{lem:d1-la for b_n}
	For every~$n\in\Nat$, the unary language
	$\lang{L}_n=\set{a^{2^n(n+1)}}$ is recognized by
	\begin{itemize}[nosep]
		\item a \twdfacg with $\bigO(n)$ states and $3$ annotation symbols;
		\item a \dla with $\bigO(n)$ states and $3$ work alphabets.
	\end{itemize}
\end{lemma}%
\begin{proof}
	The case~$n=0$ is trivial, so we suppose~$n>0$.
	According to~\cref{lem:2dfa for b_n},
	there exists a $\Gamma$\=/predictive \twdfa~$\machine{A}$ over~$\Gamma=\set{0,1,\ssep}$
	with~$\bigO(n)$ states
	which recognizes~$\set{b_n}$.
	Since~$\lang{L}_n=\norm{\set{b_n}}$,
	the first statement directly follows from \cref{prop:2nfa+cg for norm},
	and the second statement follows from \cref{prop:2dfa to d1-la}.
\end{proof}

In order to prove that every~\twnfa recognizing~$\lang{L}_n$
requires more than~$2^n$ states,
we use the following lemma,
which is implied by \cite[Theorem 3.5]{MP01}.%
\begin{lemma}
	\label{lem:2nfa loop}
	If an $n$\=/state \twnfa accepts a word $a^m$ for some $m\geq n$,
	then there exist~$C\geq1$ and~$N\geq0$
	such that it accepts the word~$a^{m+kC}$ for every~$k\geq N$.
	In particular, it accepts infinitely many words.
\end{lemma}

As a consequence, $2^n(n+1)+1$ states are required (and actually sufficient)
for a \twnfa to recognize the \emph{singleton}~$\lang{L}_n$.
Hence, in combination with \cref{lem:d1-la for b_n,lem:2dfa+cg for b_n},
we obtain the following result which in particular improves~\cite[Theorem~2]{PP19}.%
\begin{theorem}
	\label{thm:gap d1-la -> 2nfa}
	\label{thm:gap 2dfa+cg -> 2nfa}
	For every $n\in\Nat$,
	there exists a unary language~$\lang{L}_n$ such that:
	\begin{enumerate}[nosep]
		\item $\lang{L}_n$ is recognized by a \twdfacg with~$\bigO(n)$ states and~$3$ annotation symbols;
		\item $\lang{L}_n$ is recognized by a \dla with~$\bigO(n)$ states and~$3$ work symbols;
		\item Each \twnfa requires $2^n(n+1)+1$ states to recognize~$\lang{L}_n$.
	\end{enumerate}
\end{theorem}

\section{Exponential gap from \twdfacgs to \dlas and \twnfas}
\label{sec:2dfa+cg to d1-la}

In this section, we analyze the size cost for converting \twdfacg into \owdfa, \twnfa, and \dla.
We first present an upper bound for simulating \twdfacgs using \ownfas and \owdfas,
followed by the exposition of the corresponding lower bounds,
which are obtained from a unary language.

\begin{proposition}
	Every $n$\=/state \twdfacg has an equivalent
	\begin{enumerate}
		\item\label{item:2dfa+cg -> 1nfa} $(n+1)^{n+1}$-state \ownfa;
		\item\label{item:2dfa+cg -> 1dfa} $2^{2^{(n+1)\log{(n+1)}}}$-state \owdfa.
	\end{enumerate}
\end{proposition}%
\begin{proof}
	Let $\braket{\machine{A},\Sigma,\Gamma}$ be a \twdfacg with $n$ states.
	First, we eliminate the two-wayness from the \twdfa~\machine{A},
	by applying the Shepherdson's construction (see~\cite{She59}).
	This conversion yields a \owdfa~\machine{A'} with $(n+1)^{n+1}$ states over~$\Sigma\times\Gamma$.

	The next step is transforming $\braket{\machine{A'},\Sigma,\Gamma}$ into a \ownfa~\machine{B} over~$\Sigma$.
	This can be achieved by preserving the state set of \machine{A'},
	and defining the transitions as follows:
	for each state $q$ and symbol $\sigma\in\Sigma$, the set of states accessible from $q$ in the \ownfa through $\sigma$,
	corresponds to the set of states reachable from $q$ in \machine{A'}, via
	transitions labeled by $(\sigma,\gamma)$ for some $\gamma\in\Gamma$,
	\ie, denoting by~$\delta_{\machine{B}}$ and~$\delta_{\machine{A}'}$
	the transition functions of~$\machine{B}$ and~$\machine{A}'$, respectively,
	\[
		\delta_{\machine{B}}(q,\sigma)=\bigcup_{\gamma\in\Gamma}\delta_{\machine{A}'}(q,(\sigma,\gamma))
		\text.
	\]

	This construction proves~\cref{item:2dfa+cg -> 1nfa}.
	To prove~\cref{item:2dfa+cg -> 1dfa},
	it is possible to obtain an equivalent~\owdfa by applying the powerset construction to~\machine{B}.
\end{proof}

We now introduce a family of unary languages $\left(\lang{M}_n\right)_{n\geq1}$,
each of which is succinctly represented by a \twdfacg but requires a double exponential
number of states to be recognized by a \owdfa.
To define these languages,
given an integer~$n\geq1$,
first consider the language \defem{\ifbs} over the alphabet~$\set{0,1,\ssep,\dsep}$
composed by repetitions of some non-empty suffix of $b_n$ followed by a marker~$\dsep$.
Formally:
\[
	\ifbs = \set{(u\dsep)^k | k\geq 1,\; u\in\suff{b_n},\; u\neq\emptyword}\text.
\]
The length of each word $(u\dsep)^k$ in this set is clearly divisible by
$\lvert u\dsep\rvert=\lvert u\rvert+1$,
which is greater than $1$ and less than or equal to~$\lvert b_n\rvert+1$.
Thus, the length of every word in this language has a divisor between~$2$ and~$2^n(n+1)+1$.
The unary language~$\lang{M}_n$ is defined as the language of lengths of~$\ifbs$, \ie,%
\begin{align*}
	\lang{M}_n	&= \set{a^{kd}\mid k\geq0,\;1<d\leq 2^n(n+1)+1}
	\\	&= \norm{\ifbs}
	\text.
\end{align*}

\subsection{A small \twdfacg for~$\ifbs$}
By definition of~$M_n$ and \cref{prop:2nfa+cg for norm},
in order to define a \twdfacg recognizing~$M_n$,
it suffices to define a \twdfa recognizing \ifbs.
This will be obtained by first building a \twdfa of size linear in~$n$
recognizing the auxiliary language~${(\suff{b_n}\dsep)}^*$
which supersets~$\ifbs$.
To this end, we prove the following preliminary lemma.%
\begin{lemma}
	\label{lem:mark iteration of 2dfa}
	If \machine{M} is an $n$\=/state \twdfa over $\Sigma$ and $\dsep\notin\Sigma$,
	then the language ${(\langof{\machine{M}}\dsep)}^*$ is recognized by a $(2n+1)$-state \twdfa.
\end{lemma}
\ifshort%
\begin{proof}[Proof sketch]
	First, we put~$\machine{M}$ in some convenient form in which:
	\begin{itemize}[nosep]
		\item the initial state~$s$ has no incoming transition;
		\item each state other than~$s$ additionally store
			the direction~$d\in\set{-1,+1}$
			to which the head was lastly moved.
	\end{itemize}
	This can be done using~$2n+1$ states, preserving the recognized language.

	Then, we add transitions on symbol~$\dsep$,
	so that, from state~$p$ storing~$d\in\set{-1,+1}$ as above-explained,
	$\dsep$ is interpreted as the left endmarker if~$d=-1$ and as the right endmarker if~$d=+1$.
	Next, from each copy of a final state~$q_f$ storing~$d=+1$,
	we set~$\delta(q_f,\dsep)=\set{(s,+1)}$.
	Finally, we define~$s$ to be the initial and the unique final state.
	The resulting machine is a $2n+1$-state \twdfa
	recognizing~$\langof{\machine{M}\dsep}^*$.
\end{proof}
\else%
\begin{proof}
	The main idea is to repeatedly simulate~$\machine{M}$
	on each factor delimited by~$\dsep$ (or by~$\lend$ and~$\dsep$),
	treating the special symbol~$\dsep$ as an endmarker.
	According to whether a cell containing~$\dsep$ is entered from the right or from the left,
	the symbol is interpreted as the left or the right endmarker, respectively.
	To distinguish these two cases, we extend the state set, so that the direction of the last head move is saved.

	We first define an intermediate \twdfa~$\machine{M}'$ equivalent to~$\machine{M}$
	which stores the direction of the last head move
	in its finite control.
	Technically, its state set consists in two copies of the state set~$Q$ of~$\machine{M}$
	--- one for each direction of the previous head move ---,
	plus the initial state
	--- which has no preceding head move and that we keep for technical reasons.
	Hence, $\machine{M}'$ has~${2n+1}$ states.
	Formally,
	we let~%
	${Q'=\set{i}\cup\set{q_c\mid q\in Q,\;c\in\set{-1,+1}}}$,
	$F'=\set{q_{+1}\mid q\in F}$ if~$i\notin F$ or~$\set{q_{+1}\mid q\in F}\cup\set{i}$ otherwise,
	and, for each~$p\in Q$, each~$c\in\set{-1,+1}$, and each~$\sigma\in\Sigma\cup\set{\lend,\rend}$,
	${\delta'(p_c,\sigma)=(q_d,d)}$ when~${(q,d)\in\delta(p,\sigma)}$
	and~${\delta'(i,\sigma)=\delta'(i_{-1},\sigma)=\delta'(i_{+1},\sigma)}$.
	We then let~${\machine{M}'=\braket{Q',\Sigma,\delta',i,F'}}$.
	By construction, $\machine{M}'$
	is equivalent to~$\machine{M}$,
	has~$2n+1$ states,
	and its initial state~$i$ has no incoming transition.

	We now define the~$2n+1$\=/state \twdfa~${\machine{M}''=\braket{Q',\Sigma,\delta'',i,\set{i}}}$
	which recognizes~${{\left(\langof{\machine{M}}\dsep\right)}^*}$,
	by defining~$\delta''$ as follows.
	For~$p\in Q'$ and~$\sigma\in\Sigma\cup\set{\lend,\dsep,\rend}$:
	\begin{align*}
		\delta''(p,\sigma)&=
		\begin{cases}
			\delta'(p,\sigma)	&	\text{if~$\sigma\neq\dsep$ and~$(p,\sigma)\neq(i,\rend)$}														\\
			\delta'(p,\lend)	&	\text{if~$\sigma=\dsep$ and~$p\in\set{q_{-1}\mid q\in Q}$}													\\
			\delta'(p,\rend)	&	\text{if~$\sigma=\dsep$ and~$p\in(\set{q_{+1}\mid q\in Q}\cup\set{i})\setminus F'$}	\\
			(i,+1)						&	\text{if~$\sigma=\dsep$ and~$p\in F'$}																							\\
			\text{undefined}	&	\text{otherwise}
		\end{cases}
	\end{align*}
	By ensuring~$(p,\sigma)\neq(i,\rend)$ in the case~$\sigma\neq\dsep$,
	we force the machine~$\machine{M}''$ to halt as soon as it enters an accepting configuration,
	namely, when it enters state~$i$ with the head scanning~$\rend$.
	In particular, the empty input is accepted.
	Also, since~$i$ has no incoming transition in~$\machine{M}'$
	and because~$i$ is the initial state of~$\machine{M}''$,
	we have that, in every computation of~$\machine{M''}$,
	the state~$i$ is entered exactly when visiting for the first time a cell immediately to the right of a cell containing~$\lend$ or~$\dsep$.
	Moreover, for every input~$u_1\dsep u_2\dsep\cdots u_r\dsep$ with~$u_1u_2\cdots u_r\in\Sigma^*$, the following property holds:
	if, for some~$i\leq r$, the \twdfa~$\machine{M}''$ starts in state~$i$
	with the head at position~${\lvert u_1\dsep\cdots u_{i-1}\dsep\rvert}$
	(scanning the first symbol of~$u_i\dsep$),
	then it eventually enters the state~$i$ with the head at position~${\lvert u_1\dsep\cdots u_i\dsep\rvert}$
	(scanning the first symbol of~$u_{i+1}\dsep$ if~$i<r$ or~$\rend$ otherwise)
	\ifof
	$u_i\in\langof{M}$.
	This inductively leads to~$\langof{M}''={(\langof{M}\dsep)}^*$.
\end{proof}
\fi

From \cref{lem:mark iteration of 2dfa,lem:2dfa for b_n and special factors} we obtain:%
\begin{lemma}
	\label{lem:mark iteration of suffix of b_n}
	For every positive integer $n$, the language
	${(\suff{b_n}\dsep)}^*$ is recognized by a \twdfa with $\bigO(n)$ states.
\end{lemma}

Next, to recognize which word~$w\in{(\suff{b_n}\dsep)}^*$ belongs to~\ifbs,
we define a second auxiliary language,
named \defem{$\sameprefix{n}$},
that will also be recognized by a \twdfa of linear size in~$n$.
Given a positive integer~$n$%
,
let us define the language%
\begin{align*}
	\sameprefix{n}
	&=\set{(u\dsep)^k | k\geq1,u\in\set{0,1,\ssep}^*, 0<\lvert u\rvert<n} \\
	&\cup\set{ux_1\dsep\cdots ux_k\dsep | k\geq1,ux_i\in\set{0,1,\ssep}^*\text{ for~$i = 1, \ldots, k$}, \lvert u\rvert=n}\text.
\end{align*}
Each word of \sameprefix{n} can be decomposed into factors separated by a marker~\dsep,
which have the following property:
either they are shorter than~$n$ and equal,
or they are longer than~$n$ and they share the same prefix of length~$n$.

The next result states that there exists a~\twdfa that recognizes~\sameprefix{n}
with a small number of states.%
\begin{lemma}
	\label{lem:2dfa for same prefix}
	For every positive integer $n$%
	,
	the language~\sameprefix{n}
	is recognized by a~\twdfa with~$\bigO(n)$ states.
\end{lemma}%
\begin{proof}
	Let $w\in\set{0,1,\ssep,\dsep}^*$.
	A \defem{block} of $w$ is defined as a maximal factor that ends with~\dsep,
	and that does not contain any other occurrence of \dsep.
	Thus, $w$ can be decomposed into a sequence of non-overlapping blocks.

	To verify that $w\in\sameprefix{n}$, a \twdfa can check,
	for~$j = 1,\ldots, n$,
	that the symbols in position~$j$ of all blocks are equal.
	We shall treat the case in which all blocks have length less than~$n$ as a particular case.

	This verification can be performed as follows
	(refer to \cref{proc:isSameSuffix}).
	For each~$j$,
	the automaton moves to the cell in position~$j$ of the first block
	and stores the symbol contained therein
	in a variable~$\sigma$ (\cref{l:proc/isSamePrefix:store sigma}).
	Then,
	it moves the head to the $j$\=/th position of each block,
	one by one,
	to check if the symbol in that position matches the symbol stored in~$\sigma$
	(\cref{l:proc/isSamePrefix:compare sigma}).

	More precisely,
	after matching the $j$\=/th symbol of a block,
	the automaton moves the head to the right until it reaches the end of the current block
	(\ie, when it reads $\dsep$, \cref{l:proc/isSamePrefix:end of block}).
	The head is then moved~$j$ positions to the right
	to reach the $j$\=/th cell of the next block
	and check if it matches~$\sigma$.
	Once the $j$\=/th symbols of all blocks have been compared,
	the automaton moves the head to the left endmarker
	(\cref{l:proc/isSamePrefix:to lend})
	and repeats this process to compare the $(j+1)$\=/th symbols of each block.

	The \twdfa halts and accepts the input in two cases:
	\begin{itemize}
		\item the check is successfully completed for all $j = 1, \ldots, n$
			(\cref{l:proc/isSamePrefix:length>=n}), or
		\item
			the check is successfully completed for all $j = 1, \ldots, \ell$,
			for some $\ell < n$,
			and the~$\ell$-th symbol of all blocks is $\dsep$
			(\cref{l:proc/isSamePrefix:length<n}).
	\end{itemize}

	It is possible to observe that
	the size of the \twdfa implementing this procedure is linear in~$n$,
	because it uses a finite counter to store~$j$,
	whose values range from~$1$ to~$n$,
	and the variable~$\sigma$,
	which can store~$4$ different symbols (\ie, the input symbols~$0$, $1$, $\ssep$, and~\dsep).

	Notice that the \twdfa does not need to use any
	additional variable to reach the $j$\=/th position of the blocks.
	Instead, it can locally compute and recover $j$ as follows
	(in \cref{proc:isSameSuffix},
	this is exectuted by calling the macro~\readrelative).
	To reach the $j$\=/th symbol of each block,
	the machine decrements the counter storing~$j$ to $0$ while moving to the right.
	Then,
	after having read the symbol,
	before moving to the next block,
	it recovers~$j$ by moving the head to the first symbol of the block
	while incrementing the counter at each move to the left.
	Observe that the first symbol of each block is either adjacent to $\dsep$ or $\rend$,
	making its position easily recoverable.
	\begin{procedure}[tb]
		\caption{()\space
			\issameprefix{}
			\newline
			accepts if the input word belongs to \sameprefix{n}; rejects otherwise.
			The macro \readrelative{$k$},
			which is always called when the head scans a cell containing a symbol~$X\in\set{\lend,\dsep}$,
			is used to move the head~$k$ positions to the right,
			read the input symbol
			(which is returned at the end of the macro),
			and go back to~$X$.
			If
			the device tries to perform a move to the right while reading the right endmarker with
			(in \readrelative or at \cref{l:proc/isSamePrefix:end of block}),
			the input is rejected.
			}%
			\label{proc:isSameSuffix}
			\For{$j\gets1$ \KwTo~$n$}{
				\lRepeat{$\read{}=\lend$}{move the head to the left}\label{l:proc/isSamePrefix:to lend}
				$\sigma\gets\readrelative{$j$}$\;\label{l:proc/isSamePrefix:store sigma}
					\While{$\readrelative{$1$}\neq\rend$}{
						\lIf{$\readrelative{$j$}\neq\sigma$}{\reject} \label{l:proc/isSamePrefix:compare sigma}
						\lRepeat{$\read{}=\dsep$}{move the head to the right}\label{l:proc/isSamePrefix:end of block}
					}
					\lIf{$\sigma=\dsep$}{\accept}\label{l:proc/isSamePrefix:length<n}
				}
			\accept\label{l:proc/isSamePrefix:length>=n}
		\end{procedure}
\end{proof}

We are now ready to design our small \twdfa recognizing \ifbs.%
\begin{proposition}
	\label{prop:2dfa for ifbs}
	For every positive integer $n$, the language
	\[
		\ifbs=\set{(u\dsep)^k \mid k\geq1,u\in\suff{b_n}, u\neq\emptyword}
	\]
	is recognized by \twdfa with $\bigO(n)$ states.
\end{proposition}%
\begin{proof}
	First,
	observe that \cref{it:factor identify suffix} of \cref{prop:factor identify suffix} implies that
	\[
		\ifbs={(\suff{b_n}\dsep)}^*\cap\sameprefix{n+1}
		\text.
	\]
	Hence this result is a consequence of \cref{lem:mark iteration of suffix of b_n,lem:2dfa for same prefix}.
\end{proof}

\subsection{Lower bounds}
\label{sec:lower bounds from 2dfa+cg to others}
Our goal is now to show that the minimum \owdfa recognizing~$\lang{M}_n$
requires a number of states that is doubly exponential in~$n$.
In fact, the exact lower bound is expressed using the notion of \emph{primorial}.
For a positive integer~$k$ the \defem{primorial of~$k$},
denoted by~$\defem{\primorial{k}}$,
is the product of all prime numbers less than or equal to~$k$, \ie,
\[
	\primorial{k} = \prod\set{p \in \mathbb{N} \mid p \text{ is prime and } p \leq k}.
\]
The following result is used to obtain the doubly exponential lower bound.%
\begin{lemma}
	\label{lem:primorial-exp lower bound}
	For $n\geq 0$, $\primorial{(2^n(n+1)+1)}>2^{2^nn}$.
\end{lemma}%
\begin{proof}
	If~$n<4$ then the statement can be checked by simple calculations.
	We thus assume~$n\geq 4$ and we let~$k_n=2^n(n+1)+1$.
	Our goal is to prove~$\primorial{k_n}>2^{2^nn}$.
	Because~$n\geq 4$, $k_n\geq 41$.
	Using the monotony of~$\ln$,
	we get~$\frac{1}{\ln(k_n)}\leq\frac{1}{\ln(41)}<1-\ln(2)$,
	which is rewritten as:
	\begin{align*}
		1-\frac{1}{\ln(k_n)}
		&>
		\ln(2)
		\text.
	\end{align*}
	Now, as for every~$k\geq 41$,
	$\ln(\primorial{k})>k(1-\frac{1}{\ln(k)})$~\cite[Formula~(3.16)]{RS62},
	where~\defem{$\ln(x)$} denotes the natural logarithm of~$x$,
	we deduce:
	\begin{align*}
		\ln(\primorial{k_n})
		&>
		k_n\left(1-\frac{1}{\ln(k_n)}\right)
		>
		k_n\ln(2)
	\end{align*}
	which, exponentiated, yields~$\primorial{k_n} > 2^{k_n} = 2^{2^n(n+1)+1} > 2^{2^nn}$.
\end{proof}

We now recall some notions about the strong structure enjoyed by unary regular languages,
see, \eg,~\cite{Pig15}.
We say that a unary language~$\lang{L}$ over~$\set{a}$ is \defem{$C$\=/cyclic} for some~$C\geq1$,
if for every~$p\in\Nat$, either both or none of~$a^p$ and~$a^{p+C}$ belong to~$\lang{L}$.
If furthermore,
there are no~$C'<C$ such that~$\lang{L}$ is~$C'$\=/cyclic,
then it is said \defem{properly $C$\=/cyclic}.
A language is said \emph{cyclic}
if it is $C$\=/cyclic for some~$C\geq1$.
A unary language~$\lang{L}$ is regular \ifof
it is \emph{ultimately cyclic},
namely,
for some finite language~$\lang{F}$ and some cyclic language~$\lang{I}$,
$\lang{L}=\lang{F}\cdot\lang{I}$.
Also, if~$\lang{L}$ is properly $C$\=/cyclic,
then the minimum \owdfa recognizing it has~$C$ states.%
\begin{lemma}
	\label{lem:M_n properly cyclic}
	\label{lem:1dfa for M_n}
	Let~$n\geq 0$ and~$k_n=2^n(n+1)+1$.
	The language~$\lang{M}_n$ is properly~$(\primorial{k_n})$\=/cyclic
	and, consequently, recognized by a minimum $(\primorial{k_n})$\=/state \owdfa.
\end{lemma}%
\begin{proof}
	It is routine to prove that,
	for every~$\ell\in\Nat$,
	$a^{\ell}\in\lang{M}_n$ \ifof $a^{\ell+\primorial{k_n}}\in\lang{M}_n$.
	Hence, $\lang{M}_n$ is $(\primorial{k_n})$\=/cyclic.

	Let~$C$ be such that~$\lang{M}_n$ is $C$\=/cyclic.
	Our goal is to show that~$C\geq\primorial{k_n}$.
	It is sufficient to prove that every prime number~$p\leq k_n$ divides~$C$.
	Let~$p$ be a prime that does not divide~$C$.
	By Dirichlet's theorem on primes in arithmetic progressions (see, \eg, \cite{Bat04}),
	there exists some~$i>k_n$ such that~$p+iC$ is prime.
	Since~$p+iC$ is prime and larger than~$k_n$,
	it follows that $a^{p+iC}\notin\lang{M}_n$.
	This implies that $a^{p}\notin\lang{M}_n$,
	and consequently~$p\nleq k_n$.
	Hence, for every prime~$p$,
	if~$p\leq k_n$ then~$p$ divides~$C$.
	Thus $C\geq\primorial{k_n}$.
\end{proof}

We are now ready to prove the main theorem of the paper:%
\begin{theorem}
	\label{thm:2dfa+cg for M_n}%
	\label{thm:1dfa for M_n}%
	Let~$n\in\Nat$ and $\lang{M}_n$ be the unary language
	\[
		\lang{M}_n=\set{a^{kd}\mid k\geq0,\; 1<d\leq 2^n(n+1)+1}
		\text.
	\]
	Then:
	\begin{enumerate}[nosep]
		\item $\lang{M}_n$ is recognized by a \twdfacg with~$\bigO(n)$ states and~$4$ annotation symbols.%
		\item The minimum \owdfa recognizing $\lang{M}_n$ has~$\primorial{(2^n(n+1)+1)}>2^{2^nn}$ states.
	\end{enumerate}
\end{theorem}%
\begin{proof}
	For~$n=0$ the statement holds trivially.
	So assume~$n\geq 1$ and let~$k_n=2^n(n+1)+1$.

	Since $\lvert b_n\rvert=2^n(n+1)=k_n-1$, for every natural number~$d$, we have
	$1<d\leq k_n$ \ifof there exists a non-empty suffix~$u$ of~$b_n$ such that~$\lvert u\dsep\rvert=d$.
	Thus, it follows that ${\lang{M}_n=\norm{\ifbs}}$.
	By applying \cref{prop:2nfa+cg for norm,prop:2dfa for ifbs},
	we conclude that~$\lang{M}_n$ is recognized by a \twdfacg
	with~$\bigO(n)$ states and annotation alphabet~${\Gamma=\set{0,1,\ssep,\dsep}}$.
	The second statement of the theorem is directly given by \cref{lem:M_n properly cyclic,lem:primorial-exp lower bound}.
\end{proof}

As a consequence of~\cref{thm:2dfa+cg for M_n},
we obtain the following gaps for the other models studied in this paper.
In particular, we solve~\cite[Problem~1]{Pig19}\xspace%
about the size cost of the simulation of \la by \owdfa in the unary case.
Moreover, by exhibiting an exponential lower bound for the simulation of \la by \dla in the unary case,
we contribute to the investigation of~\cite[Problem~2]{Pig19}
which asks for the size cost of determinizing \las.
Whether a single exponential is always sufficient for this conversion is still an open problem,
in the general and in the unary case
(the best-known upper bound being doubly-exponential).%
\begin{theorem}
	\label{cor:d1-la for M_n}
	For every $n\in\Nat$,
	there exists a unary language $\lang{L}_n$ such that:
	\begin{enumerate}[nosep]
		\item\label{statement:2dfa+cg for M_n} A~\twdfacg with~$\bigO(n)$ states and~$\bigO(1)$ annotation symbols recognizes it;
		\item\label{statement:1-la for M_n} A~\la with~$\bigO(n)$ states and~$\bigO(1)$ work symbols recognizes it;
		\item\label{statement:d1-la for M_n} Every \dla needs at least~$2^n$ states to recognize it;
		\item\label{statement:2nfa for M_n} Every \twnfa needs more than~$2^n$ states to recognize it.
	\end{enumerate}
\end{theorem}%
\begin{proof}
	Let $\lang{L}_n=\lang{M}_n$.
	By \cref{thm:2dfa+cg for M_n},
	the \hyperref[statement:2dfa+cg for M_n]{first item} holds,
	and the \hyperref[statement:1-la for M_n]{second item}
	is obtained by simulating the~\twdfacg for~$\lang{L}_n$ with a \la
	(see \cref{sec:prelim} and~\cite{GP19} for further details).

	For the \hyperref[statement:d1-la for M_n]{third item},
	the lower bound for \dlas is a consequence of the upper bound of $m(m+1)^m$ states
	for the conversion an~$m$-state \dla into a~\owdfa~\cite{PP14},
	and the fact that the minimal \owdfa for~$\lang{L}_n$ has more than~$2^{{2^n}n}$ states.
	Suppose that there exists an $m$\=/state \dla
	recognizing~$\lang{L}_n=\lang{M}_n$
	with~$m<2^n$.
	Then, by the conversion of \dlas into \owdfas,
	the size of the obtained \owdfa would be~%
	$m(m+1)^m<(m+1)^{m+1}\leq(2^n)^{2^n}=2^{n2^n}$,
	which contradicts \cref{thm:1dfa for M_n}.

	In order to prove the \hyperref[statement:2nfa for M_n]{fourth item},
	we again proceed by contradiction
	and assume that there exist an $m$\=/state \twnfa recognizing~$\lang{L}_n$ with~$m\leq2^n$.
	By \cite[Theorem~3.6]{MP01},
	there exists~$\ell\geq1$ such that
	for all sufficiently large~$p$ and all~$k\geq0$,
	$a^p\in\lang{L}_n$ \ifof $a^{p+k\ell}\in\lang{L}_n$.
	Furthermore, $\ell$ can be found smaller than or equal to~$F(m)$,
	where~$F(m)$ is known in the literature as \emph{the Landau's function}.%
	\footnote{The explicit upper bound~$F(m)$ for~$ell$ is given in the proof of \cite[Theorem~3.6]{MP01}.}
	On the one hand,
	as shown in~\cite{Mas84},
	for every~$m\geq1$,
	${\ln(F(m)) \leq 1.05313\sqrt{m \ln(m)}}$,
	which implies that~${F(m)<2^m}$ since ${1.05313\sqrt{m\ln(m)}<m\ln(2)}$.
	Hence, with~$m\leq2^n$, we have:%
	\footnote{\Cref{eq:period upper bound} holds when~$n>0$; for~$n=0$, the statement of the theorem is trivially satisfied.}
	\begin{equation}
		\label{eq:period upper bound}
		\ell\leq F(m)<2^m\leq2^{2^n}<2^{2^nn}
		\text.
	\end{equation}
	On the other hand,
	by \cref{lem:M_n properly cyclic,lem:primorial-exp lower bound}:
	\begin{align}
		\label{eq:period lower bound}
		\ell&\geq\primorial{(2^n(n+1)+1)}>2^{2^nn}
		\text.
	\end{align}
	We conclude by observing that \cref{eq:period lower bound,eq:period upper bound} are in contradiction.
\end{proof}
\section{Complement of~$\lang{M}_n$}
\label{sec:complement}
As shown in \hyperref[section: annotation to d1-la]{the previous section},
the language~$\lang{M}_n=\set{a^{kd}\mid k\geq0,\; 1<d\leq 2^n(n+1)+1}$
is recognized by a \twdfacg (which is a nondeterministic device) of size linear in~$n$ (\cref{thm:2dfa+cg for M_n})
but requires an exponential size at least to be recognized by one of the deterministic devices considered so far,
in particular~\dla and~\twdfa (\cref{cor:d1-la for M_n}).
A natural question when dealing with nondeterministic devices
is the size cost arising from complementation.
Indeed, though the cost of complementing deterministic devices is usually cheap,%
\footnote{%
	See~\cite{GMP07,GP19}
	for the complementations of~\twdfa and \dla
	with linear and polynomial size cost,
	respectively.%
}
it is a more intricate task when starting from a nondeterministic device.
In particular the question of the cost of complementing \twnfa is open.
In fact, showing a super-polynomial lower bound for this conversion,
in combination with the linear size cost upper bound of the corresponding question for~\twdfa~\cite{GMP07},
would imply Sakoda \& Sipser conjecture.

In this section we prove an exponential size gap
for the conversion of a~\la into another one recognizing the complement of the language.
This is obtained as a consequence of a doubly-exponential lower bound
for the size of a~\ownfa recognizing the complement~${a}^*\setminus\lang{M}_n$ of~$\lang{M}_n$.
Hence the lower bounds is obtained even in the special case of a unary language.

We actually give two proofs of the doubly-exponential lower bound for the size of a \ownfa recognizing the complement of~$\lang{M}_n$.
The first one uses the standard \emph{extended fooling set} technique.
The latter, which gives a stronger bound, rely on a result from~\cite{MP05},
and shows that the smallest \ownfa recognizing~${a}^*\setminus\lang{M}_n$
has as many state as the equivalent minimum~\owdfa
(which has more than~$2^{2^nn}$ states by~\cref{lem:1dfa for M_n,lem:primorial-exp lower bound}).

\subparagraph{First proof using the extended fooling set technique.}
Using the standard \emph{extended fooling set} technique~\cite{Bir92},
we prove that~$2^{2^{n-1}}$ states are required for a~\ownfa to recognize~${a}^*\setminus\lang{M}_n$.
Notice that this implies the same lower bound for a~\owdfa to recognize~$\lang{M}_n$,
that is,
a lower bound having the same order as those obtained with a direct proof in \cref{thm:1dfa for M_n},
but slightly weaker.%
\begin{lemma}
	\label{lem:1nfa for a*-M_n using efs}
	For each~$n\geq1$,
	every \ownfa recognizing~${a}^*\setminus\lang{M}_n$
	has at least~$2^{2^{n-1}}$ states.
\end{lemma}%
\begin{proof}
	The lower bound is obtained by the well-known \emph{extended fooling set} technique.\xspace
	An \emph{extended fooling set} for a language~$\lang{L}$
	is a set~$F=\set{(x_0,y_0), (x_1,y_1), \ldots, (x_\ell,y_\ell)}$ of pairs of words
	such that,
	on the one hand~$x_iy_i\in\lang{L}$ for every~$i=0,\ldots,\ell$,
	and on the other hand $x_iy_j\notin\lang{L}$ or~$x_jy_i\notin\lang{L}$ for~$0\leq j < i \leq \ell$.
	Given an extended fooling set~$F$ of a regular language~$\lang{L}$,
	every \ownfa recognizing~$\lang{L}$ has at least~$\card{F}$ states~\cite[Lemma~1]{Bir92}.

	Let~$P_n$ be the set of prime numbers that are smaller than or equal to~$2^n(n+1)+1$.
	For each~$X\subseteq P_n$,
	we let~$\compl{X}=P_n\setminus X$ and $d_X=\prod X$ (with the natural convention that~$d_\emptyset=1$).
	Also, for each subset~$X$ of~$P_n$,
	we observe that the unary word of length~$d_X+d_{\compl{X}}$
	does not belong to~$\lang{M}_n$.
	Indeed, every prime number less than or equal to~$2^n(n+1)+1$ divides exactly one of~$d_X$ and~$d_{\compl{X}}$,
	and thus not their sum.

	Let~$X$ and~$Y$ be two distinct subsets of~$P_n$.
	Then, at least one of the unary words of length~$d_X+d_{\compl{Y}}$ and~$d_Y+d_{\compl{X}}$ belongs to~$\lang{M}_n$.
	Indeed, since~$X$ and~$Y$ are different,
	there exists,
	without loss of generality,
	a prime~$p\in X\setminus Y$.
	It is possible to notice that~$p$ divides both~$d_X$ and~$d_{\compl{Y}}$, and hence~$d_X+d_{\compl{Y}}$.
	Therefore,
	$\set{(a^{d_X},a^{d_{\compl{X}}}) \mid X\subseteq P_n}$ is an extended fooling set for~${a}^*\setminus\lang{M}_n$.
	Consequently, every \ownfa accepting the language has at least~$2^{\card{P_n}}$ states.

	Now we prove that,
	for every~$n\geq 1$,
	the inequality~$\card{P_n}>2^{n-1}$ holds.
	We treat the cases~$n=1$ and~$n=2$ separately.
	We have $\card{P_1}=3$ and~$\card{P_2}=6$,
	thus the inequality holds in both cases.
	We now assume~$n\geq 3$, which, in particular, implies~$2^n(n+1)+1\geq 17$.
	Since the number of primes less than a constant~$k$
	is greater than~$\frac{k}{\ln(k)}$ for~$k\geq17$ \mbox{\cite[Corollary~1]{RS62}},
	we get:
	\begin{align*}
		\card{P_n}	&>\frac{2^n(n+1)+1}{\ln(2^n(n+1)+1)}>\frac{2^n(n+1)}{2(n+1)}=2^{n-1}
	\end{align*}
	This concludes the proof.
\end{proof}

\subparagraph{Second proof using~\cite{MP05}.}
In~\cite{MP05},
the authors showed that,
if a unary language~$\lang{L}$
is recognized by a \emph{``small''} \ownfa
then every \ownfa recognizing its complement
has as many states as its minimum~\owdfa.
Here, ``small'' means that the language witnesses
the maximal state gap for the conversion of \ownfas into \owdfas.
For~$C$\=/cyclic languages (\cf \cref{sec:lower bounds from 2dfa+cg to others})
with prime factorization~$C=p_1^{r_1}\times\cdots\times p_s^{r_s}$
(the~$p_i$'s are distinct prime numbers, and the $r_i$'s are positive integers),
this is implied by the existence of a \ownfa
with no more than~$p_1^{r_1}+\cdots+p_s^{r_s}$ many states in cycles~\cite[Theorem~3.1]{MP05}.%
\footnote{In~\cite[Theorem~3.1]{MP05}, the result is stated in the more general case of \emph{ultimately} $C$\=/cyclic languages.}
Now,
by \cref{lem:M_n properly cyclic},
we have that~$\lang{M}_n$ is $\primorial{k_n}$\=/cyclic where~${k_n=2^n(n+1)+1}$.
Remember that, by definition, ${\primorial{k_n}=\prod\set{p\in\mathbb{N} \mid p \text{ is prime and } p \leq k_n}}$.
Hence, a ``small'' \ownfa for~$\lang{M}_n$ should have $\defem{\pi(k_n)}=\sum\set{p\in\mathbb{N} \mid p \text{ is prime and } p \leq k_n}$ states in cycles.
\begin{lemma}
	\label{lem:small 1nfa for M_n}
	\label{lem:large 1nfa for a*-M_n}
	\label{lem:1nfa for a*-M_n using MP05}
	For each~$n\geq 1$,
	\begin{enumerate}
		\item there exists a \ownfa recognizing~$\lang{M}_n$ with~$\pi(2^n(n\!+\!1)\!+\!1)\!+\!1$ states,
			whose initial state has no incoming transition whence does not participate to any cycle;%
		\item each \ownfa recognizing~${a}^*\!\setminus\!\lang{M}_n$
			has at least $\primorial{(2^n(n\!+\!1)\!+\!1)}$
			(whence more than~$2^{2^nn}$) 
			states.%
	\end{enumerate}
\end{lemma}
\begin{proof}
	Let~$n\geq1$ and let~$k_n=2^n(n+1)+1$.
	For each~$p\geq1$, the language~$\set{a^p}^*$ is recognized by a $p$\=/state \owdfa.
	Therefore, the first statement directly follows from the following identity:
	$\lang{M}_n = \bigcup_{p\in{P}_n}\set{a^p}^*$
	where~$P_n$ is the set of primes less than or equal to~$k_n$.
	Then, the second statement is a direct consequence of the first one by \cite[Theorem~3.1]{MP05}.%
\end{proof}

\subparagraph{Lower-bound for complementing unary \las.}
Since the conversion of an $n$\=/state~\la into an equivalent~\ownfa is known to cost at most~$n2^{n^2}$ in size~\cite{PP14},
it follows
that every \la recognizing~${a}^*\setminus\lang{M}_n$
requires an exponential number of states in~$n$.
This contrasts with the linear size of the \la (actually, the \twdfacg) recognizing~$\lang{M}_n$ given in~\cref{thm:2dfa+cg for M_n}.%
\begin{theorem}
	\label{thm:1-la for a*-M_n}
	Let~$n\in\Nat$ and $\lang{M}_n$ be the unary language
	\[
		\lang{M}_n=\set{a^{kd}\mid k\geq0,\; 1<d\leq 2^n(n+1)+1}
		\text.
	\]
	Then:
	\begin{enumerate}[nosep]
		\item\label{it:2dfa+cg for M_n again}$\lang{M}_n$ is recognized by a \twdfacg with~$\bigO(n)$ states and~$4$ annotation symbols.%
		\item\label{it:1-la for a*-M_n}Every~\la needs more than~$2^{\frac{n}{2}}$ states to recognize~${a}^*\setminus\lang{M}_n$.
	\end{enumerate}
\end{theorem}%
\begin{proof}
	The \hyperref[it:2dfa+cg for M_n again]{first item} is given by \cref{thm:2dfa+cg for M_n}.
	In order to prove the \hyperref[it:1-la for a*-M_n]{second item},
	let~$\machine{M}$ be a~\la recognizing~${a}^*\setminus\lang{M}_n$
	and let~$m$ denote its number of states.
	By~\cite[Theorem~2]{PP14},
	there exists a \ownfa recognizing~${a}^*\setminus\lang{M}_n$
	with no more than~$m\cdot 2^{m^2}$ states.
	Proceeding by contradiction, we assume~$m\leq 2^{\frac{n}{2}}$.
	Then:
	\ifshort%
	\(
		m\cdot2^{m^2}	\leq 2^{\frac{n}{2}} \cdot 2^{2^{n}}<2^{2^n}
		\text.
	\)
	\else%
	\[
		m\cdot2^{m^2}	\leq 2^{\frac{n}{2}} \cdot 2^{2^{n}}<2^{2^n}%
		\text.
	\]%
	\fi%
	This contradicts \cref{lem:1nfa for a*-M_n using MP05} which states that~$m\cdot 2^{m^2}\geq2^{2^{n}n}\geq 2^{2^n}$.
	Hence, $m>2^{\frac{n}{2}}$.
\end{proof}

\bibliographystyle{plainurl}
\bibliography{references}

\begin{thebibliography}{10}

\bibitem{Bat04}
Paul~Trevier Bateman and Harold~G Diamond.
\newblock {\em Analytic number theory: an introductory course}, volume~1.
\newblock World Scientific, 2004.

\bibitem{Bir92}
Jean-Camille Birget.
\newblock Intersection and union of regular languages and state complexity.
\newblock {\em Information Processing Letters}, 43(4):185--190, 1992.

\bibitem{BDGP17}
Miko{\l}aj Boja{\'n}czyk, Laure Daviaud, Bruno Guillon, and Vincent Penelle.
\newblock Which classes of origin graphs are generated by transducers?
\newblock In {\em {ICALP} 2017}, volume~80 of {\em LIPIcs}, pages
  114:1--114:13. Schloss Dagstuhl - Leibniz-Zentrum f{\"{u}}r Informatik, 2017.

\bibitem{GMP07}
Viliam Geffert, Carlo Mereghetti, and Giovanni Pighizzini.
\newblock Complementing two-way finite automata.
\newblock {\em Information and Computation}, 205(8):1173--1187, 2007.
\newblock \href {https://doi.org/10.1016/j.ic.2007.01.008}
  {\path{doi:10.1016/j.ic.2007.01.008}}.

\bibitem{GP19}
Bruno Guillon and Luca Prigioniero.
\newblock Linear-time limited automata.
\newblock {\em Theoretical Computer Science}, 798:95--108, 2019.

\bibitem{HU79}
John~E Hopcroft and Jeffrey~D Ullman.
\newblock {\em Introduction to Automata Theory, Languages and Computation}.
\newblock Addison-Wesley, 1979.

\bibitem{Mas84}
Jean-Pierre Massias.
\newblock Majoration explicite de l'ordre maximum d'un {\'e}l{\'e}ment du
  groupe sym{\'e}trique.
\newblock {\em Annales de la Facult{\'e} des sciences de Toulouse :
  Math{\'e}matiques}, 5e s{\'e}rie, 6(3-4):269--281, 1984.
\newblock URL: \url{https://www.numdam.org/item/AFST_1984_5_6_3-4_269_0/}.

\bibitem{MP05}
Filippo Mera and Giovanni Pighizzini.
\newblock Complementing unary nondeterministic automata.
\newblock {\em Theoretical Computer Science}, 330(2):349--360, 2005.
\newblock \href {https://doi.org/10.1016/j.tcs.2004.04.015}
  {\path{doi:10.1016/j.tcs.2004.04.015}}.

\bibitem{MP01}
Carlo Mereghetti and Giovanni Pighizzini.
\newblock Optimal simulations between unary automata.
\newblock {\em {SIAM} J. Comput.}, 30(6):1976--1992, 2001.
\newblock \href {https://doi.org/10.1137/S009753979935431X}
  {\path{doi:10.1137/S009753979935431X}}.

\bibitem{Pig15}
Giovanni Pighizzini.
\newblock Investigations on automata and languages over a unary alphabet.
\newblock {\em International Journal of Foundations of Computer Science},
  26(07):827--850, 2015.

\bibitem{Pig19}
Giovanni Pighizzini.
\newblock Limited automata: Properties, complexity and variants.
\newblock In {\em {DCFS} 2019}, volume 11612 of {\em Lecture Notes in Computer
  Science}, pages 57--73. Springer, 2019.

\bibitem{PP14}
Giovanni Pighizzini and Andrea Pisoni.
\newblock Limited automata and regular languages.
\newblock {\em International Journal of Foundations of Computer Science},
  25(07):897--916, 2014.

\bibitem{PP19}
Giovanni Pighizzini and Luca Prigioniero.
\newblock Limited automata and unary languages.
\newblock {\em Inf. Comput.}, 266:60--74, 2019.

\bibitem{PP23a}
Giovanni Pighizzini and Luca Prigioniero.
\newblock Forgetting 1-limited automata.
\newblock {\em Electronic Proceedings in Theoretical Computer Science},
  388:95--109, 2023.
\newblock \href {https://doi.org/10.4204/eptcs.388.10}
  {\path{doi:10.4204/eptcs.388.10}}.

\bibitem{PP23b}
Giovanni Pighizzini and Luca Prigioniero.
\newblock Once-marking and always-marking 1-limited automata.
\newblock {\em International Journal of Foundations of Computer Science},
  (Online ready):1--24, 2025.
\newblock \href {https://doi.org/10.1142/S0129054124440015}
  {\path{doi:10.1142/S0129054124440015}}.

\bibitem{RS62}
J~Barkley Rosser and Lowell Schoenfeld.
\newblock Approximate formulas for some functions of prime numbers.
\newblock {\em Illinois Journal of Mathematics}, 6(1):64--94, 1962.

\bibitem{SS78}
William~J. Sakoda and Michael Sipser.
\newblock Nondeterminism and the size of two way finite automata.
\newblock In {\em {STOC} 1978}, pages 275--286. {ACM}, 1978.
\newblock \href {https://doi.org/10.1145/800133.804357}
  {\path{doi:10.1145/800133.804357}}.

\bibitem{She59}
John~C Shepherdson.
\newblock The reduction of two-way automata to one-way automata.
\newblock {\em IBM Journal of Research and Development}, 3(2):198--200 
  0018--8646, 1959.

\bibitem{Wag89}
Klaus~W Wagner and Gerd Wechsung.
\newblock {\em Computational Complexity}.
\newblock D. Reidel Publishing Company, Dordrecht, 1986.

\end{thebibliography}
\end{document}